\documentclass{Odyssey2026}

 \interspeechcameraready

\title{Similarity Choice and Negative Scaling in Supervised Contrastive Learning for Deepfake Audio Detection}

\author[]{Jaskirat}{Sudan}
\author[]{Hashim}{Ali}
\author[]{Surya}{Subramani}
\author[]{Hafiz}{Malik}

\affiliation{CECS}{University of Michigan, Dearborn}{USA}

\email{jsudan@umich.edu, alhashim@umich.edu, suryasss@umich.edu, hafiz@umich.edu}

\usepackage{comment}
\usepackage{soul}
\begin{document}

\maketitle

\begin{abstract}

Supervised contrastive learning (SupCon) is widely used to shape representations, but has seen limited targeted study for audio deepfake detection. Existing work typically combines contrastive terms with broader pipelines; however, the focus on SupCon itself is missing. In this work, we run a controlled study on wav2vec2 XLS-R (300M) that varies (i) similarity in SupCon (cosine vs angular similarity derived from the hyperspherical angle) and (ii) negative scaling using a warm-started global cross-batch queue. Stage 1 fine-tunes the encoder and projection head with SupCon; Stage 2 freezes them and trains a linear classifier with BCE. Trained on ASVspoof 2019 LA and evaluated on ASV19 eval plus ITW and ASVspoof 2021 DF/LA, Cosine SupCon with a delayed queue achieves the best ITW EER (8.29\%) and pooled EER (4.44), while angular similarity performs strongly without queued negatives (ITW 8.70), indicating reduced reliance on large negative sets.

\end{abstract}

\section{Introduction}

Recent advances in neural text-to-speech (TTS) and voice conversion (VC) systems have drastically improved the realism of synthetic speech. These modern generative systems are capable of synthesizing speech that is perceptually indistinguishable from genuine recordings. While these technologies enable beneficial applications including personalization, accessibility, and multilingual dubbing, they also pose significant security risks. Contemporary voice cloning systems can replicate a target speaker from only a few seconds of audio, facilitating fraud, impersonation, and misinformation attacks. The rapid improvement and accessibility of speech synthesis systems make robust deepfake audio detection an urgent and evolving challenge.

To mitigate these threats, speech anti-spoofing has evolved from handcrafted features to neural detectors trained end-to-end, with recent systems increasingly leveraging large self-supervised speech encoders as front-end feature extractors combined with lightweight back-end classifiers.\cite{yi2024add2022audiodeep,tak2022automaticspeakerverificationspoofing} While SSL-based pipelines substantially improve in-domain performance on benchmarks such as ASVspoof~\cite{todisco2019asvspoof,yamagishi2021asvspoof}, their accuracy often degrades under distribution shifts, particularly in cross-dataset and in-the-wild (ITW) conditions \cite{ali2024audio,ali2026superb}. Therefore, improving generalization to unseen synthesis methods and recording environments remains a central challenge in deepfake audio detection.

One promising direction to address this OOD degradation is to explicitly regularize the representation space using contrastive objectives. Recent anti-spoofing systems incorporate supervised contrastive learning either by combining it with classification losses on top of SSL embeddings or by constructing stronger contrastive batches via augmentation and re-synthesis \cite{tran2024spoofed,doan_2024_trident_of_poseidon,doan_2024_triad,wu2024cladrobustaudiodeepfake}. These results suggest that contrastive training can improve robustness, but key design choices in SupCon are often adopted by default rather than studied in isolation.


Supervised contrastive learning (SupCon)~\cite{khosla2021supervisedcontrastivelearning} 
forms positive pairs from samples sharing the same label and negatives 
from samples with different labels, encouraging intra-class compactness 
and inter-class separation. Two design choices that directly shape what 
SupCon learns are largely fixed by default in prior work. First, the 
\emph{similarity function}: cosine similarity (equivalently, the dot 
product on $\ell_2$-normalized embeddings) is the standard choice, 
implicitly fixing the geometry of pairwise comparisons. However, the 
gradient of cosine similarity with respect to the inter-vector angle 
vanishes for very similar or very dissimilar pairs, which may affect 
optimization dynamics and temperature sensitivity. Second, the 
\emph{negative set size}: contrastive objectives benefit from many 
negatives~\cite{khosla2021supervisedcontrastivelearning,chen2020simpleframeworkcontrastivelearning,he2020momentumcontrastunsupervisedvisual,oord2019representationlearningcontrastivepredictive,robinson2021contrastivelearninghardnegative}, but large-batch training 
is impractical for SSL speech encoders due to GPU memory constraints. 
Neither choice has been systematically studied for deepfake audio 
detection.


Motivated by these gaps, we present a controlled study of two SupCon 
design choices for cross-dataset deepfake speech detection using 
wav2vec2 XLS-R~(300M) as a fixed backbone. While prior work treats 
similarity and negative scaling as fixed implementation details, we 
argue that these choices are consequential design decisions that 
interact with each other and with temperature in non-obvious ways. 
Holding all other factors constant like architecture, pooling, 
optimizer, augmentation, and Stage~2 classifier, we isolate the 
effect of each axis independently, enabling a cleaner attribution of 
performance differences to the contrastive design choices rather than 
to confounding pipeline factors. To our knowledge, the interaction 
between similarity geometry and negative scaling has not been 
previously characterized in the deepfake audio detection literature, 
and the practical implications of this interaction for cross-dataset 
generalization remain unexplored. We keep the SupCon objective 
unchanged and study two design axes:
\begin{enumerate}
    \item \textbf{Similarity function:} we replace the default cosine 
    similarity in SupCon with an angular (geodesic) similarity computed 
    from the normalized dot product, and evaluate its impact on OOD EER 
    across a temperature sweep;
    \item \textbf{Negative scheduling:} we scale the number of 
    negatives via a global cross-batch memory queue, enabled only after 
    an initial warm-up to mitigate early representation drift, and 
    study how the interaction between queue size and similarity choice 
    affects cross-dataset generalization.
\end{enumerate}

\begin{figure*}[t]
  \centering
  \includegraphics[width=\linewidth]{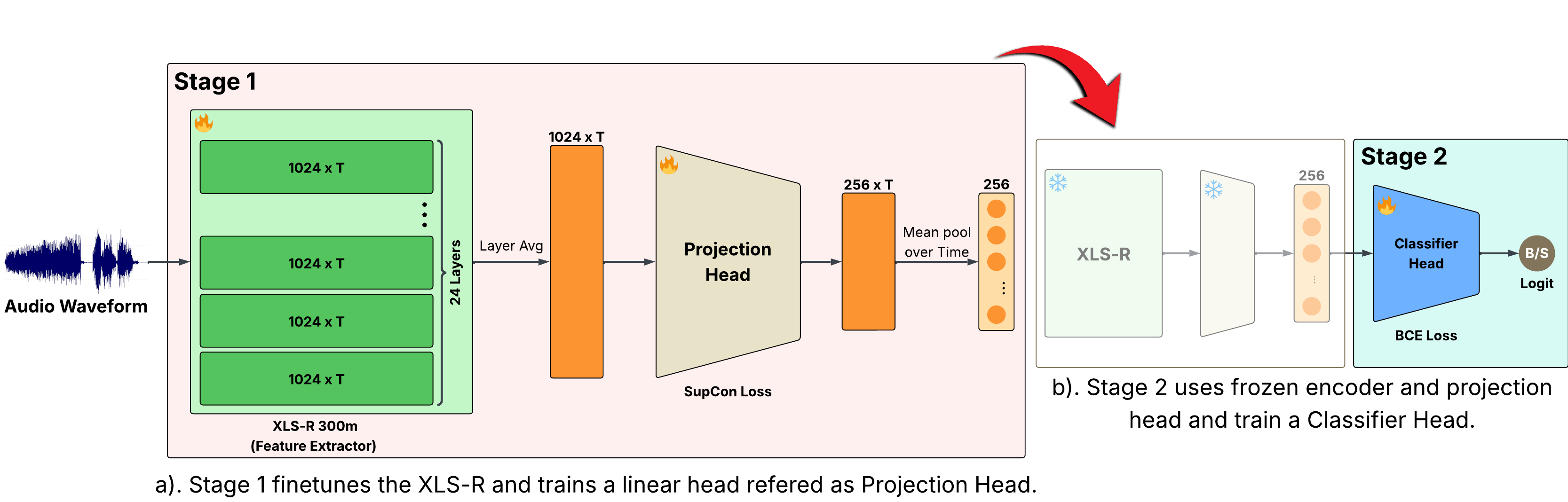}
  \caption{Model Architecture}
  \label{fig:model_architecture}
\end{figure*}

\section{Related Work}
\label{sec:related_work}

Recent work has used supervised contrastive learning (SupCon) to improve robustness, but typically by \emph{engineering hard negatives} and \emph{structuring mini-batches}. Trident of Poseidon introduces a triad training strategy that combines supervised contrastive learning with hard negative mining through audio re-synthesis and proactive batch sampling to balance bona fide/spoof composition within each batch \cite{doan_2024_trident_of_poseidon}. Balance/augmentation/re-synthesis training strategies similarly assemble mini-batches around anchors, augmentations, and re-synthesized samples, and customize the SupCon objective to reflect that batch construction \cite{doan_2024_triad}. While effective, these approaches tightly couple the contrastive signal to the availability and fidelity of re-synthesized hard negatives and to per-batch sampling design, and can bias learning toward vocoder-style deepfakes.

Another line of work applies SupCon directly to embeddings extracted from large self-supervised speech encoders. Speaker-embedding style pipelines based on WavLM combine layer-wise selection and pooling with a joint cross-entropy and supervised contrastive objective, reporting that the contrastive term stabilizes performance across evaluation sets \cite{tran2024spoofed}. These results support contrastive regularization as a useful tool for spoof detection, but they do not isolate how negative \emph{scale} and \emph{timing} influence training stability and cross-dataset performance.

Overall, prior SupCon-based anti-spoofing methods emphasize \emph{which negatives to construct} (e.g., re-synthesis) and \emph{how to form batches} (e.g., balancing and multi-view sampling), but leave under-explored the question of \emph{how many negatives to use and when to introduce them} when operating under realistic batch-size and compute constraints. Moreover, these works largely adopt the default inner-product family for contrastive similarity either explicit cosine similarity or temperature-scaled dot products in the logits \cite{doan_2024_trident_of_poseidon,doan_2024_triad,tran2024spoofed,wu2024cladrobustaudiodeepfake} without investigating whether the similarity geometry itself affects generalization. In this work, we address both gaps by (i) introducing a \emph{delayed cross-batch negatives curriculum} that progressively expands the negative set via a \emph{delayed cross-batch memory queue} to improve stability, and (ii) evaluating a \emph{hyperspherical geodesic (angular)} similarity on $\ell_2$-normalized embeddings as an alternative angular similarity parameterization to cosine/dot-product similarity.

\section{Method}
\label{sec:method}

We study supervised contrastive learning for deepfake speech detection
in a controlled two-stage XLS-R pipeline. The encoder architecture,
projection dimension, pooling strategy, Stage~2 classifier, optimizer,
and augmentation policy are fixed across all experiments. We vary only
two factors in Stage~1: the similarity function used inside SupCon
(Section~\ref{subsec:similarity}) and the scale and timing of
cross-batch negatives (Section~\ref{subsec:queue_negatives}).

\subsection{Pipeline overview}
\label{subsec:pipeline}

An XLS-R encoder produces a layer-aggregated frame sequence, which a
linear projection head maps to a compact utterance embedding via
temporal mean pooling and $\ell_2$-normalization, yielding
$\tilde{\mathbf{z}}$. Architecture details and exact dimensions are
given in Section~\ref{subsec:model_setup}.

Training proceeds in two stages. In Stage~1, the encoder and projection
head are fine-tuned jointly using the SupCon objective applied to
$\tilde{\mathbf{z}}$. In Stage~2, both are frozen and a linear
classifier is trained on the fixed embeddings using binary
cross-entropy. All variants share this pipeline; only the Stage~1
objective and negative set differ.

\subsection{Supervised contrastive objective}
\label{subsec:supcon}

We adopt supervised contrastive learning (SupCon) to learn embeddings
that are compact within class and separated across classes. For each
anchor example in a minibatch, samples sharing the same label
(bona-fide or spoof) are treated as positives, while samples from the
opposite label are treated as negatives. SupCon encourages the anchor
to be closer to its positives than to all other candidates by
maximizing a temperature-scaled log-softmax probability assigned to
positives.

For a minibatch $\mathcal{B}$, let $P(i)$ denote indices with the same
label as anchor $i$ (excluding $i$), and let $A(i)$ denote the
candidate set used in the denominator. The SupCon loss is:

\begin{equation}
\mathcal{L}_{\mathrm{supcon}}
=
\sum_{i\in \mathcal{B}}
\frac{-1}{|P(i)|}
\sum_{p \in P(i)}
\log
\frac{
\exp(\mathrm{sim}(\tilde{\mathbf{z}}_i,\tilde{\mathbf{z}}_p)/\tau)
}{
\sum_{a \in A(i)}
\exp(\mathrm{sim}(\tilde{\mathbf{z}}_i,\tilde{\mathbf{z}}_a)/\tau)
}
\label{eq:supcon}
\end{equation}

The temperature $\tau > 0$ controls the concentration of the 
softmax distribution, and its optimal value depends on the 
similarity function used~(Section~\ref{sec:results}).

We consider two candidate set configurations. In \emph{batch-only}
SupCon, $A(i)$ contains all non-self samples in the current minibatch.
In \emph{queue-augmented} SupCon, $A(i)$ is extended with negatives
drawn from a cross-batch memory queue
(Section~\ref{subsec:queue_negatives}). In both cases, positives
$P(i)$ are drawn from the current minibatch only; the queue contributes
additional negatives exclusively. This restricts stale embeddings to
the denominator, avoiding reliance on stale same-class structure while
increasing negative diversity.

\subsection{Similarity: cosine vs.\ geodesic}
\label{subsec:similarity}

Supervised contrastive learning is commonly implemented using cosine
similarity on $\ell_2$-normalized
embeddings~\cite{khosla2021supervisedcontrastivelearning}. Under
normalization, representations lie on a unit hypersphere, where cosine
similarity depends only on the inter-vector angle. We compare two
SupCon variants that differ only in the similarity function:
(i)~cosine similarity,
$\mathrm{sim}_{\cos}(\tilde{\mathbf{z}}_i,\tilde{\mathbf{z}}_j)
  =\tilde{\mathbf{z}}_i^\top \tilde{\mathbf{z}}_j$,
and (ii)~an angular similarity based on the hyperspherical angle
$\theta_{ij}=\arccos(\mathrm{clip}(\tilde{\mathbf{z}}_i^\top\tilde{\mathbf{z}}_j,-1,1))$,
mapped to the cosine range as
$\mathrm{sim}_{\mathrm{geo}}(\tilde{\mathbf{z}}_i,\tilde{\mathbf{z}}_j)
= 2\left(1-\frac{\theta_{ij}}{\pi}\right)-1$.

The linear rescaling maps $\theta_{ij} \in [0, \pi]$ to $[-1, 1]$, matching the range of cosine similarity exactly: identical vectors ($\theta_{ij}{=}0$) map to $+1$ and opposite vectors ($\theta_{ij}{=}\pi$) map to $-1$. This ensures that $\mathrm{sim}_{\mathrm{geo}}$ and $\mathrm{sim}_{\mathrm{cos}}$ share the same $[-1, 1]$ range, making the temperature $\tau$ directly comparable across both variants. Without this rescaling, differences in optimal $\tau$ could reflect a scale mismatch rather than a genuine difference in similarity geometry.

We treat the geodesic form as an alternative angular similarity
parameterization inside SupCon rather than a fundamentally different
representation space. Because $\mathrm{sim}_{\mathrm{geo}}$ is a
monotonic transform of the inter-vector angle on normalized embeddings,
it induces identical pairwise rankings to cosine similarity. Their
practical difference lies in gradient behavior: since
$\mathrm{sim}_{\cos} = \cos\theta_{ij}$, its gradient with respect to
$\theta_{ij}$ is $-\sin(\theta_{ij})$, which vanishes near
$\theta_{ij} \approx 0$ and $\theta_{ij} \approx \pi$ and peaks at
$\theta_{ij} = \pi/2$. In contrast, $\mathrm{sim}_{\mathrm{geo}}$ is
linear in $\theta_{ij}$, producing a constant gradient magnitude across
all angular separations. This difference alters logit scaling and the
sensitivity of the loss to temperature $\tau$, which we study
empirically in Section~\ref{sec:results}. The clipping in the
$\arccos(\cdot)$ mapping ensures numerical stability, and the linear
rescaling to $[-1,1]$ keeps the similarity range comparable to cosine
for a controlled comparison under the same $\tau$.

\begin{figure*}
  \centering
  \includegraphics[width=\linewidth]{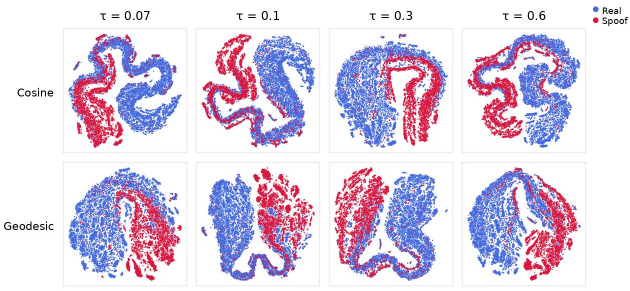}
  \caption{t-SNE visualization of ITW embeddings across temperatures (no queue).
  Rows correspond to cosine and geodesic similarity; columns to $\tau \in \{0.07, 0.1, 0.3, 0.6\}$.}
  \label{fig:temperature_sweep}
\end{figure*}

\subsection{Scaling negatives with a delayed global queue}
\label{subsec:queue_negatives}

Contrastive objectives benefit from a large and diverse negative
set~\cite{khosla2021supervisedcontrastivelearning,
he2020momentumcontrastunsupervisedvisual,
chen2020simpleframeworkcontrastivelearning}, but scaling the global
batch size is impractical for SSL speech encoders due to GPU memory
constraints. We instead reuse embeddings from previous minibatches via
a global FIFO queue $\mathcal{Q}$~\cite{wang2020crossbatchmemoryembeddinglearning}.

A notable constraint of binary SupCon in this domain is that all spoof
samples are treated as positives for each other, regardless of the
synthesis method. Spoof speech is highly heterogeneous TTS, voice
conversion, and replay attacks occupy different regions of the acoustic
space~\cite{todisco2019asvspoof,muller_does_2022} and collapsing them into a single
class may limit the discriminative resolution of the learned embeddings.
We return to this point in Section~\ref{sec:discussion}.

We maintain $\mathcal{Q}$ as a global FIFO structure that stores
utterance-level embeddings and labels gathered across GPUs. At each
iteration, current minibatch embeddings are enqueued and the oldest
entries are dequeued once capacity $|\mathcal{Q}|$ is reached. Queued
embeddings are detached from the computation graph and do not receive
gradients.

For an anchor $i$ with label $y_i$, queue negatives are defined as
$\mathcal{N}_{\mathcal{Q}}(i)=\{q\in\mathcal{Q} : y_q \neq y_i\}$,
and the candidate set is expanded as
$A(i) \leftarrow A(i) \cup \mathcal{N}_{\mathcal{Q}}(i)$.
The SupCon loss in Eq.~\ref{eq:supcon} is otherwise unchanged.

A key issue is \emph{representation drift}: early in training, cached
embeddings can be inconsistent with the current
encoder~\cite{wang2020crossbatchmemoryembeddinglearning,
kim2024gradientaccumulationmethoddense}. We adopt a two-phase schedule,
training with batch-only SupCon for the first $E_{\text{start}}$ epochs
before enabling the queue. With a global batch size of 32 and
$E_{\text{start}}{=}6$, a queue of size $|\mathcal{Q}|{=}2048$ holds
embeddings from the preceding 64 minibatches, corresponding to roughly
3--4 training epochs of history at our dataset size. Larger queues
therefore include embeddings from significantly earlier in training,
increasing the risk of stale or inconsistent negatives. Since we do not
use a momentum encoder to stabilize keys, this staleness-diversity
tradeoff is a practical constraint. We ablate $|\mathcal{Q}|$ in
Section~\ref{sec:results}; a systematic sweep over $E_{\text{start}}$
is left for future work.

\begin{figure*}[t]
  \centering
  \includegraphics[width=\linewidth]{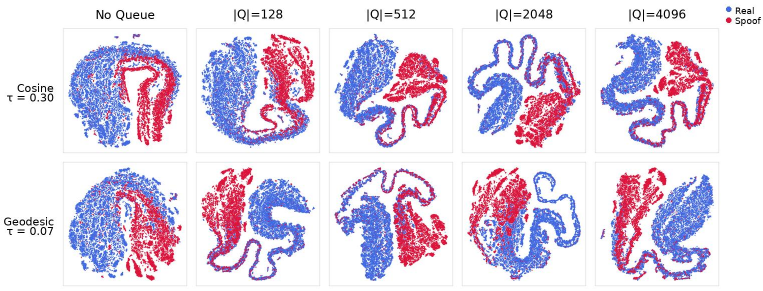}
  \caption{t-SNE visualization of ITW embeddings across queue sizes.
  Rows correspond to cosine ($\tau{=}0.30$) and geodesic ($\tau{=}0.07$);
  columns to $|\mathcal{Q}| \in \{0, 256, 1024, 4096\}$.}
  \label{fig:queue_ablation}
\end{figure*}

\subsection{Classifier training}
\label{subsec:classifier}

After Stage~1, the SSL encoder and projection head are frozen. A single
linear layer (256~$\to$~1) is trained on the fixed 256-dimensional
utterance embeddings using binary cross-entropy with logits. No
contrastive objective is used in Stage~2. This design decouples
representation shaping (Stage~1) from decision-boundary fitting
(Stage~2), allowing downstream performance differences to be attributed
primarily to the learned embedding space while keeping Stage~2 training
computationally inexpensive.

\section{Experimental Setup}
\label{sec:experiment_setup}

\subsection{Datasets and evaluation protocol}
\label{subsec:datasets}

We train all models on the ASVspoof 2019 Logical Access (LA) 
\emph{train} split and select checkpoints using the official \emph{dev} 
split~\cite{todisco2019asvspoof}. No target-domain data from ASVspoof 
2021 or ITW is used during training or model selection. In-domain 
performance is reported on ASVspoof 2019 LA \emph{eval}. To assess 
cross-dataset generalization, we additionally evaluate on the 
In-the-Wild (ITW) deepfake benchmark~\cite{muller_does_2022} and the 
official ASVspoof 2021 evaluation sets for DeepFake (DF) and 
LA~\cite{yamagishi2021asvspoof}.

ASVspoof 2019 LA eval provides near-in-domain performance, ASVspoof 
2021 DF and LA introduce codec and channel conditions unseen during 
training, and ITW contains real-world deepfakes spanning a broad range 
of synthesis systems, together covering a spectrum from near-in-domain 
to fully out-of-distribution conditions.

\subsection{Preprocessing and augmentation}
\label{subsec:preprocess}

All audio is resampled to 16~kHz and converted to mono. During 
training, utterances are represented as 10~s waveform chunks: sequences 
longer than 10~s are randomly cropped to 10~s, and shorter sequences 
are zero-padded to 10~s. At evaluation, the same fixed 10~s chunking 
policy is applied (truncation only, no padding beyond the utterance 
boundary), and a single chunk per utterance is scored. During training 
only, we apply RawBoost augmentation with probability 0.7 using the 
default noise configuration from Tak et al.~\cite{tak2022rawboost}.

\subsection{Model configuration}
\label{subsec:model_setup}


We use wav2vec2 XLS-R 
(300M)~\cite{babu2021xlsrselfsupervisedcrosslingualspeech} as the 
frontend SSL encoder. Hidden states from all 24 transformer layers are 
averaged uniformly, yielding a layer-aggregated frame sequence of 
dimension $1024 \times T$. A frame-wise linear projection (1024 to 256) 
followed by temporal mean pooling produces a 256-dimensional 
utterance-level embedding, which is $\ell_2$-normalized before the 
contrastive objective or classifier is applied (Fig.~\ref{fig:model_architecture}).

In Stage~1, the XLS-R encoder and projection head are fine-tuned 
jointly using the SupCon objective applied to the normalized 
256-dimensional embedding. In Stage~2, both are frozen and a linear 
classifier (256 to 1) is trained on the fixed utterance embeddings 
using binary cross-entropy. All variants share identical encoder 
architecture, projection dimension, pooling strategy, and Stage~2 
classifier; only the Stage~1 training objective and negative set differ.

\subsection{Optimization and hyperparameters}
\label{subsec:training_setup}


All models are trained with AdamW. We use an encoder learning rate of 
$1\times10^{-5}$, a projection head learning rate of $5\times10^{-4}$, 
and weight decay $3\times10^{-3}$. No learning rate schedule is used; 
in order to keep the comparison between cosine and geodesic variants 
controlled, a schedule would introduce an additional variable that 
interacts differently with each similarity function's gradient dynamics. 
The global batch size is 32 across 2~GPUs. Model selection uses early 
stopping with patience~10 based on ASVspoof 2019 LA dev EER. We sweep 
the SupCon temperature $\tau \in \{0.07, 0.1, 0.3, 0.6\}$.

The encoder and projection head are frozen. The linear classifier is 
trained with AdamW using a learning rate of $5\times10^{-4}$ and weight 
decay $3\times10^{-3}$, with the same batch size and early stopping 
criterion as Stage~1. Stage~2 converges in a single epoch in most 
configurations, adding negligible compute cost.

\subsection{Cross-batch negatives and queue settings}
\label{subsec:queue_settings}

The global FIFO queue stores utterance-level 256-dimensional embeddings 
and binary labels gathered across GPUs. It contributes negatives only 
to the Stage~1 SupCon loss; Stage~2 is unaffected. Queued negatives are 
enabled after an initial warm-up of $E_{\text{start}}{=}6$ epochs to 
reduce early representation drift; this value was chosen on stability 
grounds and not tuned exhaustively. We ablate queue sizes 
$|\mathcal{Q}| \in \{0, 128, 512, 2048, 4096\}$. For the queue-size 
ablation, $\tau$ is fixed to the best no-queue value per similarity 
(cosine: $\tau{=}0.30$, geodesic: $\tau{=}0.07$); per-$|\mathcal{Q}|$ 
retuning is left for future work.

\subsection{Metrics and scoring}
\label{subsec:metrics}

We report equal error rate (EER,~\%) for all benchmarks. Scores are 
produced by the Stage~2 classifier logit (higher indicates more 
bona-fide-like), and EER is computed from bona-fide vs.\ spoof score 
distributions at the threshold where false acceptance and false 
rejection rates are equal (ASVspoof-style EER computation), evaluated 
at the utterance level.

In addition to per-benchmark EER, we report a \emph{pooled EER} defined 
as the arithmetic mean of ASVspoof 2019 LA eval, ITW, ASVspoof 2021 DF, 
and ASVspoof 2021 LA EERs.

\subsection{Compared methods}
\label{subsec:comparisons}

All methods use the same backbone and head dimensions: XLS-R features
are mapped by a linear projection (1024 to 256), mean-pooled and
$\ell_2$-normalized to produce a 256-dimensional utterance embedding,
and passed to a linear classifier (256 to 1). The methods differ
only in their Stage~1 training procedure:

\begin{enumerate}
\renewcommand{\theenumi}{\roman{enumi}}
\renewcommand{\labelenumi}{(\theenumi)}

    \item \textbf{BCE baseline (end-to-end):} XLS-R, projection head,
    and classifier are fine-tuned jointly using BCEWithLogits in a
    single stage. This serves as the primary reference point for all
    SupCon variants.

    \item \textbf{SupCon (cosine or geodesic):} Stage~1 fine-tunes
    XLS-R and projection head using SupCon with either cosine or
    geodesic similarity; Stage~2 freezes both and trains the classifier
    with BCE. We sweep $\tau \in \{0.07, 0.1, 0.3, 0.6\}$ for each
    similarity variant and select the best no-queue temperature per
    similarity for subsequent queue experiments (cosine:
    $\tau{=}0.30$, geodesic: $\tau{=}0.07$).

    \item \textbf{Queue-assisted SupCon:} identical to~(ii) but
    augments Stage~1 negatives with a delayed global FIFO queue.
    The queue is enabled after $E_{\text{start}}{=}6$ epochs to
    reduce early representation drift, and contributes negatives
    only to the SupCon denominator positives are always drawn
    from the current minibatch. We ablate queue sizes
    $|\mathcal{Q}| \in \{0, 128, 512, 2048, 4096\}$ with $\tau$
    fixed to the best no-queue value per similarity.

\end{enumerate}

All experiments are run on 2$\times$NVIDIA A40 GPUs. To verify
result stability, key configurations were evaluated across multiple
random seeds; observed variance was low across seeds, and we report
results for seed 1337 throughout.

\section{Results}
\label{sec:results}

To isolate the effect of the similarity geometry and the temperature $\tau$, we first perform a temperature sweep for cosine- and geodesic-based SupCon (Table~\ref{tab:temp_sweep}). 
Relative to the BCE baseline (pooled EER $7.27$), SupCon improves cross-dataset performance when $\tau$ is tuned, but the optimal $\tau$ differs by similarity.
For cosine, $\tau=0.30$ yields the best pooled EER ($5.78$), largely driven by gains on ITW and ASV21 (ITW: $12.18$ vs.\ $9.99$; ASV21-DF: $9.12$ vs.\ $6.58$; ASV21-LA: $7.54$ vs.\ $6.18$).
For geodesic similarity, a smaller temperature is preferred: $\tau=0.07$ achieves the best pooled EER overall ($5.31$) and the lowest OOD EERs across ITW/ASV21 (ITW: $12.18$ vs.\ $8.70$; ASV21-DF: $9.12$ vs.\ $6.16$; ASV21-LA: $7.54$ vs.\ $6.11$), at the cost of a small degradation on in-domain ASV19-LA ($0.23$ vs.\ $0.25$).

Based on this sweep, we fix $\tau=0.30$ for cosine and $\tau=0.07$ for geodesic in the remaining queue experiments.

\begin{figure}[t]
  \centering
  \includegraphics[width=\linewidth]{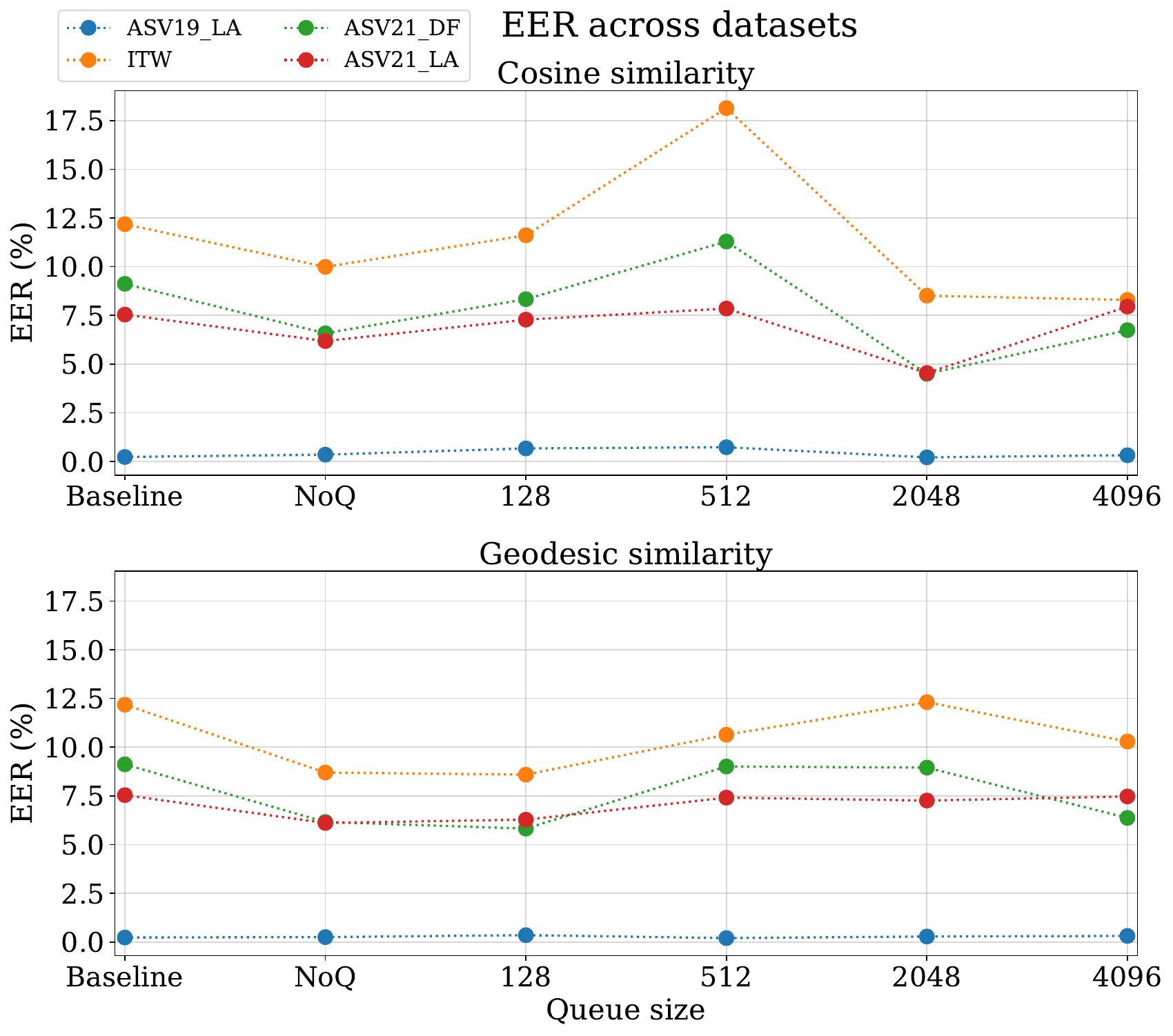}
  \caption{EER across datasets for cosine vs. geodesic and different queue sizes.}
  \label{fig:clustered_eer}
\end{figure}

We next fix the best temperatures from Table~\ref{tab:temp_sweep} (cosine $\tau{=}0.30$, geodesic $\tau{=}0.07$) and vary the number of queued cross-batch negatives (Fig.~\ref{fig:clustered_eer}). The effect of scaling negatives is strongly non-monotonic and depends on the similarity geometry. 

For cosine, small queues degrade performance (e.g., $|\mathcal{Q}|{=}128$ gives ITW $11.61$ and $|\mathcal{Q}|{=}512$ yields ITW $18.14$), but a larger queue yields substantial cross-dataset gains: at $|\mathcal{Q}|{=}2048$, EER improves on all OOD sets compared to no-queue training (ITW $9.99\to 8.51$, ASV21-DF $6.58\to 4.50$, ASV21-LA $6.18\to 4.54$), and also improves ASV19-LA ($0.35\to 0.21$). Increasing the queue further to $|\mathcal{Q}|{=}4096$ maintains a low ITW EER ($8.29$) but degrades ASV21 (DF $6.74$, LA $7.95$), indicating that simply increasing queued negatives does not guarantee better generalization.

In contrast, geodesic similarity is already strong without a queue (ITW $8.70$, ASV21-DF $6.16$, ASV21-LA $6.11$), and adding a small queue provides at best marginal improvement (e.g., $|\mathcal{Q}|{=}128$ gives ITW $8.59$ and ASV21-DF $5.82$), while larger queues consistently hurt (e.g., $|\mathcal{Q}|{=}2048$ gives ITW $12.31$). Across all queue settings, cosine with $|\mathcal{Q}|{=}2048$ achieves the lowest pooled EER (4.44), while cosine with $|\mathcal{Q}|{=}4096$ achieves the lowest ITW EER (8.29). In contrast, geodesic performs strongly without queued negatives (ITW 8.70 at $|\mathcal{Q}|{=}0$), indicating reduced reliance on large queues in the no-queue regime. These results show that the optimal temperature depends on the similarity choice, and that scaling negatives with a delayed queue yields non-monotonic behavior under fixed $\tau$ (per-$|\mathcal{Q}|$ tuning is left for future work).

\begin{table}
\centering
\small
\setlength{\tabcolsep}{4pt}      
\renewcommand{\arraystretch}{1.05}
\resizebox{\linewidth}{!}{%
\begin{tabular}{l c c c c c c}
\hline
Loss Variant & Temp. & ASV19 LA & ITW & ASV21 DF & ASV21 LA & Pooled EER \\
\hline
Baseline  & --   & 0.23 & 12.18 & 9.12 & 7.54 & 7.2675 \\
Cosine    & 0.07 & \textbf{0.21} & 11.78 & 7.88 & 6.43 & 6.575 \\
Cosine    & 0.10 & 0.29 & 14.86 & 9.38 & 6.95 & 7.87  \\
Cosine    & 0.30 & 0.35 &  9.99 & 6.58 & 6.18 & 5.775 \\
Cosine    & 0.60 & 0.43 & 14.61 & 9.00 & 6.74 & 7.695 \\
\hline
Geodesic  & 0.07 & 0.25 &  \textbf{8.70} & \textbf{6.16} & \textbf{6.11} & \textbf{5.305} \\
Geodesic  & 0.10 & 0.43 & 10.88 & 6.72 & 5.99 & 6.005 \\
Geodesic  & 0.30 & 0.32 & 14.05 & 8.54 & 6.36 & 7.3175 \\
Geodesic  & 0.60 & 1.26 & 12.51 & 10.07 & 7.43 & 7.8175 \\
\hline
\end{tabular}%
}
\caption{Temperature sweep on ASVspoof 2019 LA training.}
\label{tab:temp_sweep}
\end{table}

\section{Discussion}
\label{sec:discussion}

\subsection{Effect of temperature and similarity choice}
\label{subsec:discuss_temperature}

Table~\ref{tab:temp_sweep} shows that the optimal temperature differs
substantially between the two similarity functions: cosine peaks at
$\tau{=}0.30$ while geodesic peaks at $\tau{=}0.07$, a fourfold
difference. This is consistent with the gradient dynamics argument in
Section~\ref{subsec:similarity}: geodesic's linear dependence on
$\theta_{ij}$ makes the loss more sensitive to $\tau$, so a lower
temperature is sufficient to maintain a well-calibrated softmax
distribution. Cosine similarity requires a higher temperature to
compensate for its vanishing gradients near $\theta_{ij} \approx 0$
and $\theta_{ij} \approx \pi$.

The t-SNE panels in Fig.~\ref{fig:temperature_sweep} support this
interpretation: cosine embeddings remain mixed at $\tau{=}0.07$ and
$\tau{=}0.10$, achieving cleaner class separation only at
$\tau{=}0.30$, while geodesic embeddings already show structured
separation at $\tau{=}0.07$ and degrade at higher temperatures as the
loss becomes under-concentrated. Together, these results suggest that
the similarity choice and temperature are coupled design decisions
rather than independent hyperparameters.

\subsection{Queue ablation and embedding geometry}
\label{subsec:discuss_queue}

The EER curves in Fig.~\ref{fig:clustered_eer} show that the effect
of scaling negatives via the cross-batch queue is strongly
non-monotonic and depends on the similarity function. For cosine,
small queues hurt performance ($|\mathcal{Q}|{=}512$ gives ITW
$18.14$), but a large queue recovers and improves substantially
(see Fig.~\ref{fig:clustered_eer} and Table~\ref{tab:temp_sweep}). The
t-SNE panels in Fig.~\ref{fig:queue_ablation} reflect this: the
cosine embedding clusters become progressively more structured as
$|\mathcal{Q}|$ increases from 256 to 2048, with clearer real/spoof
separation, before showing signs of fragmentation at
$|\mathcal{Q}|{=}4096$. For geodesic, the pattern reverses
embeddings are already well-separated without a queue and cluster
structure visibly degrades as $|\mathcal{Q}|$ grows, consistent with
the rising ITW EER from $8.70$ at $|\mathcal{Q}|{=}0$ to $12.31$ at
$|\mathcal{Q}|{=}2048$.

\subsection{Why geodesic degrades with large queues}
\label{subsec:discuss_geodesic_queue}

We hypothesize that the degradation of geodesic similarity under large
queues is a consequence of its uniform gradient behavior interacting
with representation drift. Because $\mathrm{sim}_{\mathrm{geo}}$
assigns a constant gradient magnitude regardless of angular distance,
it applies equal pressure to all queued negatives including stale
embeddings from significantly earlier in training. Cosine similarity,
by contrast, has weak gradients for pairs that are already very
similar or very dissimilar, effectively down-weighting the influence
of stale embeddings that have drifted far from the current encoder's
representation space. Without a momentum encoder to stabilize the
queue keys, geodesic's uniform sensitivity to all negatives makes it
more susceptible to the inconsistency introduced by large queues. This
remains a hypothesis: a controlled experiment varying queue staleness
while holding queue size fixed would be needed to confirm it.

\subsection{Limitations}
\label{subsec:discuss_limitations}

A structural limitation of our setup is that binary SupCon treats all
spoof samples as positives for each other, regardless of the synthesis
method that produced them. In practice, TTS, voice conversion, and
replay attacks occupy different regions of the acoustic
space~\cite{todisco2019asvspoof}, and the binary label collapses this
structure. The t-SNE panels in Figs.~\ref{fig:temperature_sweep}
and~\ref{fig:queue_ablation} reflect this: even in the best
configurations, the spoof cluster is not monolithic but shows internal
dispersion, suggesting that the learned embeddings partially capture
synthesis-method variation that the contrastive objective does not
explicitly encourage. System-level or synthesis-method-level labels
would allow SupCon to separate attack types rather than collapsing
them, potentially improving both intra-class compactness and
cross-dataset generalization to unseen synthesis methods.

Beyond binary supervision, several further limitations bound the scope
of our conclusions. All experiments use a single backbone (XLS-R 300M)
trained on ASVspoof 2019 LA; whether the observed similarity and queue
effects generalize to other SSL encoders or training corpora is
unknown. The queue activation epoch $E_{\text{start}}{=}6$ was chosen
on stability grounds and not systematically tuned; different values may
alter the non-monotonic queue behavior. Finally, $\tau$ is fixed to
the best no-queue value for each similarity during the queue ablation,
meaning that some of the non-monotonic behavior may partly reflect
temperature miscalibration rather than queue size alone. Jointly
retuning $\tau$ and $|\mathcal{Q}|$ is left for future work.

\section{Conclusion and Future Work}
\label{conclusion_future_work}

We presented a controlled study of two SupCon design choices
similarity function and negative scaling for cross-dataset 
deepfake audio detection with XLS-R. Our results show that 
similarity choice and temperature are coupled: geodesic similarity 
achieves strong OOD performance at a much lower temperature than 
cosine, consistent with its uniform gradient behavior across angular 
separations. Negative scaling via a delayed cross-batch queue 
substantially benefits cosine SupCon at large queue sizes 
($|\mathcal{Q}|{=}2048$, pooled EER 4.44) but consistently 
degrades geodesic, suggesting that the optimal negative scaling 
strategy depends on the similarity function. Together, these 
findings indicate that contrastive design choices in deepfake 
detection interact in non-trivial ways and should be tuned jointly 
rather than independently.

Future work will (i) retune $\tau$ jointly with $|\mathcal{Q}|$ 
to disentangle calibration from negative scaling effects, 
(ii) incorporate drift-aware queues to mitigate staleness at 
large queue sizes, (iii) explore system-level supervision beyond 
binary labels to better capture spoof heterogeneity, and (iv) 
evaluate on more recent in-the-wild benchmarks.

\ifinterspeechfinal
     The Odyssey 2026 organisers
\else
     The authors
\fi
would like to thank ISCA and the organising committees of past Interspeech conferences for kindly providing the previous version of this template.

\bibliographystyle{IEEEtran}
\bibliography{references/paper_citations, references/refs, references/references}

\end{document}